\documentclass[a4paper,twocolumn,DIV=14]{scrartcl}

\pdfoutput=1

\usepackage[utf8]{inputenc}
\usepackage{fourier}
\usepackage[T1]{fontenc}

\usepackage{scrlayer-scrpage}
\lohead{M. Pflüger, V. Soltwisch, J. Probst, F. Scholze, M. Krumrey: GISAXS on Small Targets Using Large Beams}
\usepackage{sidecap}
\sidecaptionvpos{figure}{t}
\usepackage{afterpage}
\usepackage{graphicx}
\usepackage{amsmath}
\usepackage{amssymb}
\usepackage[exponent-product=\cdot,scientific-notation=false,separate-uncertainty=false,per-mode=symbol]{siunitx}

\usepackage{natbib}
\setcitestyle{authoryear}
\newcommand{\citeasnoun}{\citet}
\renewcommand{\cite}{\citep}

\usepackage{bm}

\newcommand{\ZZ}{\mathbb{Z}}      

\usepackage{doi}   

\usepackage[format=plain,indention=5pt,labelfont=bf,figurename=Figure,tablename=Table]{caption} 

\usepackage[protrusion=true,expansion=true]{microtype} 

\usepackage{hyperref}

\hypersetup{
    pdfauthor={Mika Pflüger, Victor Soltwisch, Jürgen Probst, Frank Scholze, Michael Krumrey},     
    pdfkeywords={GISAXS, Grazing Incidence Small Angle X-Ray Scattering, Metrology, Small Target, Gratings},
    pdftitle={GISAXS on Small Targets Using Large Beams},
    pdfsubject={The shallow incidence angles used in GISAXS lead to a very large footprint of the X-ray beam on the sample. We show that despite these large footprints, GISAXS measurements of targets with sizes down to 4 µm x 4 µm are possible.},
    colorlinks=true,       
    linkcolor=black,          
    citecolor=black,          
    filecolor=black,          
    urlcolor=blue,            
    menucolor=black,          
}

\begin{document}

\title{Grazing Incidence Small Angle X-Ray Scattering (GISAXS) on Small Targets Using Large Beams}

\author{Mika Pflüger\thanks{Contact: mika.pflueger@ptb.de} \thanks{Physikalisch-Technische Bundesanstalt (PTB), Abbestraße 2-12, 10587 Berlin, Germany} \and Victor Soltwisch\footnotemark[2] \and Jürgen Probst\thanks{Helmholtz-Zentrum Berlin (HZB), Albert-Einstein-Straße 15, 12489 Berlin, Germany} \and Frank Scholze\footnotemark[2] \and Michael Krumrey\footnotemark[2]}


\maketitle

\begin{abstract}
GISAXS is often used as a versatile tool for the contactless and destruction-free investigation of nanostructured surfaces.
However, due to the shallow incidence angles, the footprint of the X-ray beam is significantly elongated, limiting GISAXS to samples with typical target lengths of several millimetres.
For many potential applications, the production of large target areas is impractical, and the targets are surrounded by structured areas.
Because the beam footprint is larger than the targets, the surrounding structures contribute parasitic scattering, burying the target signal.
In this paper, GISAXS measurements of isolated as well as surrounded grating targets in Si substrates with line lengths from \SI{50}{\um} down to \SI{4}{\um} are presented.
For the isolated grating targets, the changes in the scattering patterns due to the reduced target length are explained.
For the surrounded grating targets, the scattering signal of a \SI{15x15}{\um} target grating structure is separated from the scattering signal of \SI{100x100}{\um} nanostructured surroundings by producing the target with a different orientation with respect to the predominant direction of the surrounding structures.
The described technique allows to apply GISAXS, e.g. for characterization of metrology fields in the semiconductor industry, where up to now it has been considered impossible to use this method due to the large beam footprint.

\end{abstract}

\section{Introduction}

For the investigation of nanostructured surfaces, grazing incidence small angle X-ray scattering (GISAXS) is now established as a powerful technique \cite{hexemer_advanced_2015,renaud_probing_2009}.
For example, GISAXS is used to investigate the active layer of solar cells ex-situ as well as in-situ \cite{gu_multi-length-scale_2012,muller-buschbaum_active_2014,rossander_situ_2014,proller_following_2016}, surface and bulk morphology of polymer films \cite{muller-buschbaum_grazing_2003,wernecke_depth-dependent_2014}, surface roughness and roughness correlations \cite{holy_x-ray_1993,holy_nonspecular_1994,babonneau_waveguiding_2009}, lithographically produced structures \cite{gollmer_fabrication_2014,soccio_morphology_2015}, and deposition growth kinetics \cite{lairson_situ_1995,renaud_real-time_2003}.	
GISAXS offers non-destructive, contact-free measurements of sample structures with feature sizes between about \SI{1}{nm} and \SI{1}{\micro\meter}, giving statistical information about the whole illuminated volume.

\begin{SCfigure*}[1.0][tb]
\includegraphics[width=1.2\columnwidth]{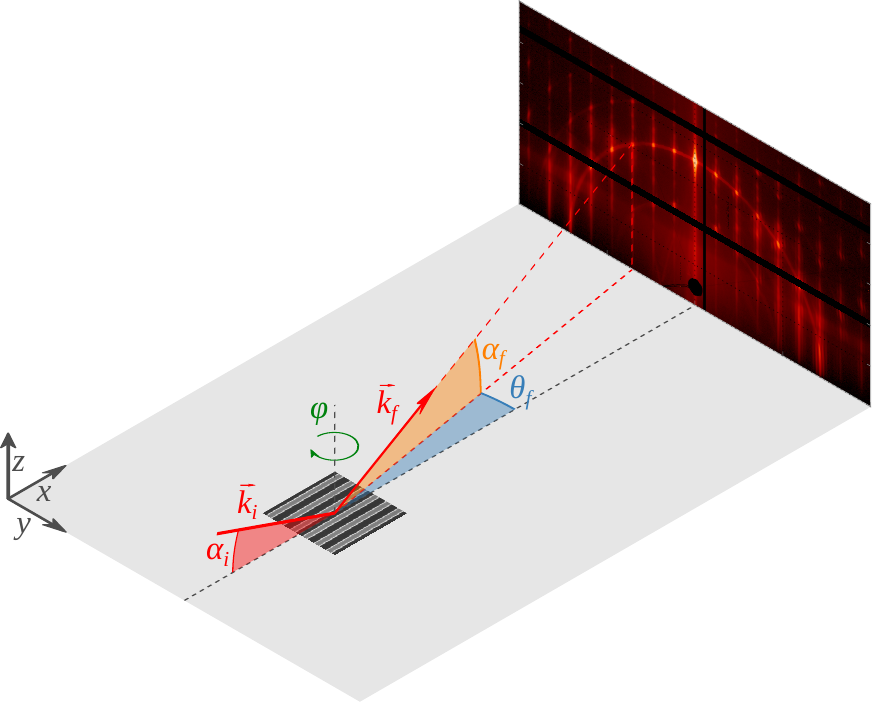}
\caption{Geometry of GISAXS experiments.
A monochromatic X-ray beam with a wavevector $\vec{k}_i$ impinges on the sample surface at a grazing incidence angle $\alpha_i$.
The elastically scattered wavevector $\vec{k}_f$ propagates along the exit angle $\alpha_f$ and the azimuthal angle $\theta_f$.
The sample can be rotated around the $z$-axis by the angle $\varphi$.}
\label{fig:gisaxs_geometry}
\end{SCfigure*}

Due to the small incidence angle $\alpha_i$ close to the angle of total external reflection $\alpha_c$ and due to the large number of scatterers in the investigated volume, scattered intensities are much higher in GISAXS geometry compared to transmission SAXS \cite{levine_grazing-incidence_1989}.
However, the low incidence angle also causes an elongated beam footprint on the sample, leading to large illuminated areas even for small incident beams.
For a typical GISAXS incidence angle of $\alpha_i \approx \SI{0.5}{\degree}$, the footprint on the sample is $\approx \num{100}$ times longer than the incident beam height.
For a moderately small beam of a synchrotron radiation beamline (height $\approx \SI{500}{\micro\meter}$), the length of the footprint on the sample is thus several centimetres.
Due to the long footprints, GISAXS has so far been routinely used only on samples which are at least several millimetres long.
To achieve shorter beam footprints, the beam height needs to be reduced.
The smallest beam height of about \SI{300}{\nano\meter} used in GISAXS experiments so far \cite{roth_situ_2007}, has led to a footprint on the sample of about \SI{30}{\micro\meter}, but presents large technical challenges in aligning the sample to the beam.
However, for many applications, the measurement of very small target areas down to a few micrometres in length is necessary, and the use of laboratory X-ray sources with comparably large beams is desirable.
A prominent application where GISAXS has been rejected so far for the mentioned reasons is the characterization of metrology fields in high-volume manufacturing of semiconductors. These fields are surrounded by other structures and larger field sizes directly translate to lost wafer area and thus additional production costs \cite{sanchez_hvm_2016}.

One approach to measuring small target areas on a surface is to use SAXS in transmission geometry.
Transmission SAXS in principle probes the whole penetrated sample volume, but it can also be used to investigate surfaces if the sample bulk is sufficiently homogeneous \cite{hu_small_2004,sunday_determining_2015}, offering a method to investigate small surface areas non-destructively and in a contact-free way.
Unfortunately, transmission SAXS is not usable for thick (with respect to the substrate material's absorption length) samples that absorb a large portion of the incoming beam nor for inhomogeneous samples where for example buried layers add to the scattering background.
For such samples, measurements in GISAXS geometry would be preferred if the problem of large illuminated areas could be overcome.

We show that GISAXS measurements of micro\-metre-sized structured surfaces are possible using existing non-focused sources for isolated targets as well as for suitably prepared periodic targets in a periodic environment.
The scattering of isolated grating targets with lengths from \SI{4}{\micro\meter} to \SI{50}{\micro\meter} is compared with the scattering of a \SI{2500}{\micro\meter} long (quasi-infinite) grating target.
We explain the length-dependent changes in the scattering patterns using the theory for slit diffraction.
For the measurement of targets surrounded by other nanostructures, we produce the grating targets with a different direction with respect to the predominant direction of their surroundings.
This allows us to separate the scattering signal of the targets from the signal of the surroundings by aligning the incident X-ray beam to the target.

\section{GISAXS at Gratings}

The measurement geometry of GISAXS \cite{levine_grazing-incidence_1989} is shown schematically in fig. \ref{fig:gisaxs_geometry}.
The sample is illuminated under grazing incidence angle $\alpha_i$, and the resulting reflected and scattered light is collected with an area detector at exit angles $\alpha_f$ and $\theta_f$.
We chose our coordinate system such that the $x$-$y$-plane is the sample plane and the $x$-axis lies in the scattering plane, with the $z$-axis perpendicular to the sample plane.
In this coordinate system, the scattering vector $\vec{q} = \vec{k}_f - \vec{k}_i$ takes the form
\begin{align}
q_x &= k (\cos \theta_f \cos \alpha_f - \cos \alpha_i)  \\
q_y &= k (\sin \theta_f \cos \alpha_f)   \\
q_z &= k (\sin \alpha_i + \sin \alpha_f)
\end{align}
with the wavevector of the incoming beam $\vec{k}_i$, the wavevector of the scattered beam $\vec{k}_f$, $k = |\vec{k}_i| = |\vec{k}_f| = 2 \pi / \lambda$ and the wavelength of the incident light $\lambda$.

Several groups have already performed GISAXS measurements on gratings, and the scattering of perfect gratings is well understood.
\citeasnoun{tolan_x-ray_1995}, \citeasnoun{metzger_nanometer_1997}, \citeasnoun{jergel_structural_1999} and \citeasnoun{mikulik_x-ray_1999} measured gratings in GISAXS geometry with the grating lines perpendicular to the incoming beam (coplanar geometry).
GISAXS measurements with the grating lines along the incoming beam (so-called non-coplanar geometry, conical mounting or sagittal diffraction geometry) were analysed by \citeasnoun{mikulik_coplanar_2001}. Their paper already contains the reciprocal space construction of the resulting scattering pattern laid out in detail by \citeasnoun{yan_intersection_2007}.
\citeasnoun{hofmann_grazing_2009} reconstructed a simple line profile using the distorted-wave Born approximation (DWBA) formalism.
\citeasnoun{hlaing_nanoimprint-induced_2011} examine the production of gratings by nanoimprinting and extract the side-wall angle of the grating profile.
For very rough polymer gratings, where the grating diffraction is not usable for the analysis, \citeasnoun{meier_situ_2012} could still extract the line profile including the side-wall angle and line width of rough polymer gratings from the diffuse part of the scattering.
Measuring rough polymer gratings as well, \citeasnoun{rueda_grazing-incidence_2012} use the DWBA formalism with form factors of different length to model gratings with varying roughness.
With a different theoretical approach, \citeasnoun{wernecke_direct_2012} and \citeasnoun{wernecke_traceable_2014} extract line and groove width as well as the line height of gratings using Fourier analysis.
Solving the Maxwell equations using finite elements, \citeasnoun{soltwisch_nanometrology_2014} and \citeasnoun{soltwisch_reconstructing_2017} reconstruct detailed line profiles of gratings, including a top and bottom corner rounding as well as the side-wall angle, the line width and height.
Most recently, \citeasnoun{suh_characterization_2016} measured rough polymer gratings and extracted the average line profile as well as the magnitude of deviations from the average line profile using DWBA.
Notably, they also showed that the reconstruction did not improve further when using a more complex line profile shape, thus demonstrating that a relatively simple line shape already describes the X-ray scattering of their grating.

The diffraction of gratings in GISAXS geometry can be described as the intersection of the reciprocal space representation of the grating and the Ewald sphere of elastic scattering \cite{mikulik_coplanar_2001,yan_intersection_2007}.
The reciprocal space representation of a grating periodically extending into infinity in the $y$-direction with infinite length and vanishing height is an array of rods (so-called grating truncation rods, GTRs) lying parallel to the reciprocal $k_y$-$k_z$-plane (see fig. \ref{fig:reciprocal_ewald_gratings} a).
The intersection of the GTRs and the Ewald sphere is a series of grating diffraction orders on a semicircle, evenly spaced in $k_y$, each $2\pi/p$ apart with the grating pitch $p$.
If the grating is rotated in the sample plane by the angle $\varphi$ such that the grating lines are no longer parallel to the $x$-axis, the GTR plane is rotated around the $k_z$-axis by $\varphi$, so that the scattering pattern becomes asymmetric.
At the small incidence angles used in GISAXS, the curvature of the Ewald sphere is very steep at the intersection, leading to large changes in the scattering pattern even for small deviations in $\varphi$ \cite{mikulik_coplanar_2001}.

Using the same construction as \citeasnoun{yan_intersection_2007}, but in the coordinate system used in this paper, the positions of the grating diffraction orders are (see supplementary material for the derivation):
\begin{align}
\alpha_f &= \arcsin\left(
\sqrt{\sin^2 \alpha_i - \left(\frac{n\lambda}{p}\right)^2 - \frac{2 n \lambda \sin \varphi \, \cos \alpha_i}{p}}
\right) \label{eq:grating_alpha_f} \\
\theta_f &= \arctan \left(
\frac{\cos \varphi \, n \lambda}{\sin \varphi \, n \lambda + p \cos \alpha_i}
\right)
\end{align}
with the X-ray wavelength $\lambda$, the grating diffraction order $n \in \ZZ$ and the grating pitch $p$.

\section{Instrumentation}

\begin{figure*}[t]
\includegraphics[width=\textwidth]{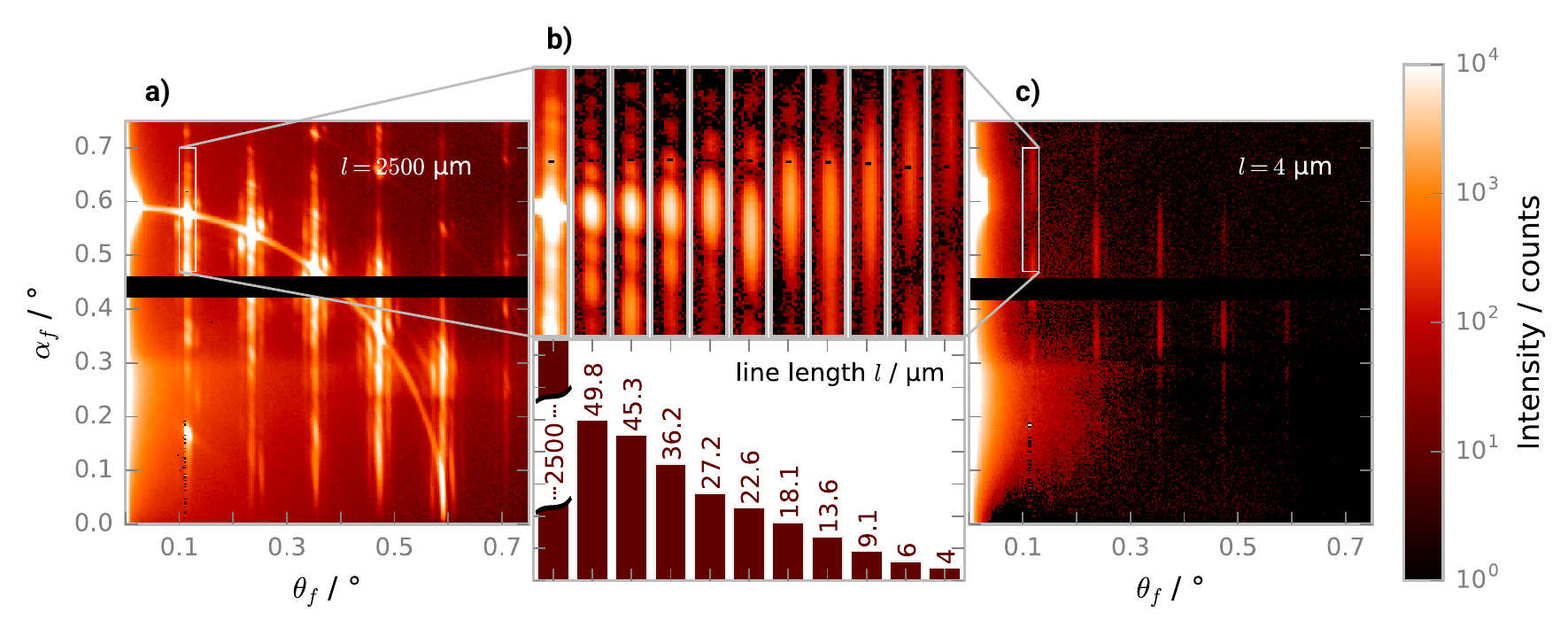}
\caption{Changes in the GISAXS pattern by line length.
\textbf{a)} GISAXS pattern of \SI{2500}{\micro\meter} long grating, showing diffraction orders on a circle.
\textbf{b)} Detailed view of the first diffraction order of gratings with differing lengths, showing the elongation of the first diffraction order with decreasing grating length (top) and corresponding grating lengths (bottom).
\textbf{c)} GISAXS pattern of \SI{4}{\micro\meter} long grating, showing the elongated diffraction orders.
For comparability, all measurements were taken with the same exposure time, which leads to overexposure for the \SI{2500}{\micro\meter} long grating.}
\label{fig:gisaxs_length_series}
\end{figure*}

\subsection{Sample Preparation}

All structures were fabricated by electron beam lithography on a Vistec EBPG5000+ using positive resist ZEP520A on silicon substrates, followed by reactive ion etching with SF$_6$ and C$_4$F$_8$ and resist stripping with an oxygen plasma treatment \cite{senn_fabrication_2011}.

\subsection{GISAXS Experiments}

The experiments were conducted at the four-crystal monochromator (FCM) beamline~\cite{krumrey_high-accuracy_2001} in the laboratory~\cite{beckhoff_quarter-century_2009} of the Physikalisch-Technische Bundesanstalt (PTB) at the electron storage ring BESSY~II. This beamline allows the adjustment of the photon energy in the range from \SI{1.75}{keV} to \SI{10}{keV}.
By using a beam-defining \SI{0.52}{mm} diameter pinhole about \SI{150}{cm} before the sample position and a scatter guard \SI{1}{mm} pinhole about \SI{10}{cm} before the sample, the beam spot size was about \SI{0.5x0.5}{mm} at the sample position with minimal parasitic scattering.
Both pinholes are low-scatter SCATEX germanium pinholes (Incoatec GmbH, Germany).
Alternatively, the beam spot size could be reduced to about \SI{0.1x0.1}{mm} by using a beam-defining \SI{100}{\micro\meter} Pt pinhole (Plano GmbH, Germany) and an adjustable slit system with low-scatter blades (XENOCS, France) as a scatter guard.
The GISAXS setup at the FCM beamline consists of a sample chamber \cite{fuchs_high_1995} and the HZB SAXS setup \cite{gleber_traceable_2010}.
The sample chamber is equipped with a goniometer which allows sample movements in all directions with a resolution of \SI{3}{\micro\meter} as well as rotations around all sample axes with an angular resolution of \ang{0.001}.
The HZB SAXS setup allows moving the in-vacuum Pilatus~1M area detector \cite{wernecke_characterization_2014}, reaching sample-to-detector distances from about \SI{2}{m} to about \SI{4.5}{m} and exit angles up to about \ang{2}.
Along the whole beam path including the sample site, high vacuum (pressure below \SI{e-4}{mbar}) is maintained.

\section{Length Series}

\begin{figure*}[t]
\includegraphics[width=\textwidth]{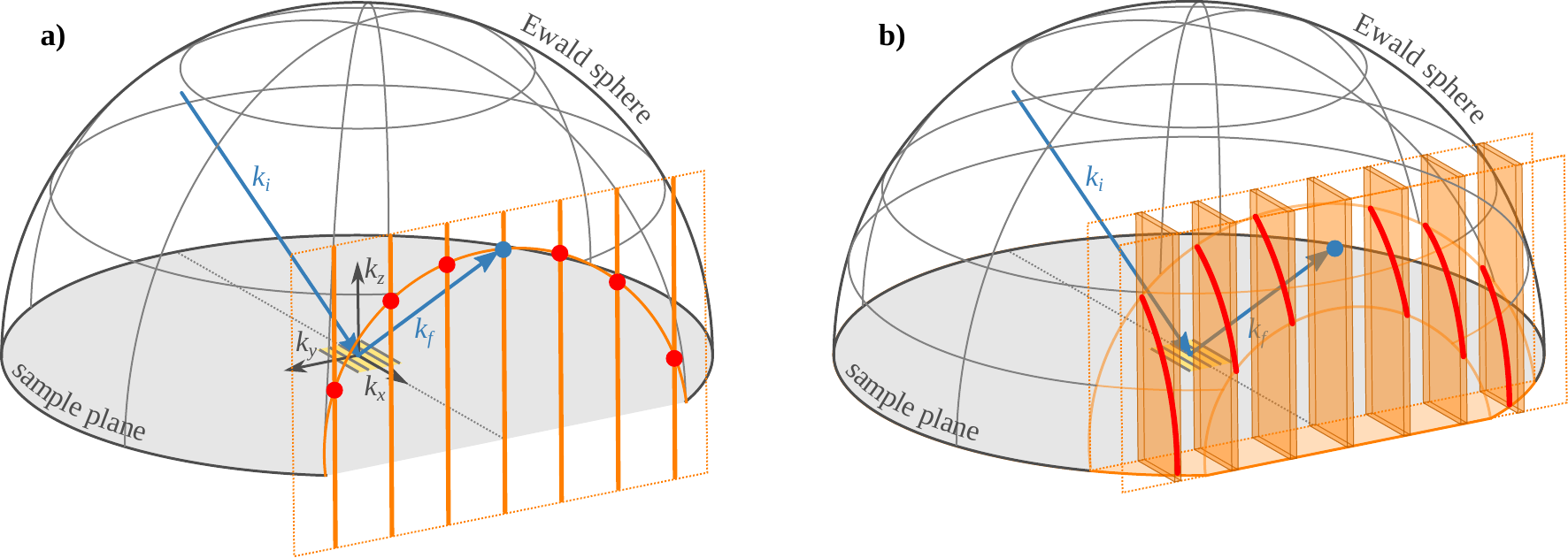}
\caption{Reciprocal space construction of GISAXS from gratings.
\textbf{a)} Ewald sphere (grey mesh) and grating truncation rods (orange), which are the reciprocal space representation of an infinite grating.
The projection of the intersection (red) on the detector (not shown) leads to the GISAXS pattern.
\textbf{b)} For a short (i.e. finitely long) grating, the reciprocal space representation (orange) along $k_x$ is not a delta function any more.
Instead, it is $\propto \mathrm{sinc} \, k_x$, leading to grating truncation sheets.
The intersection of the grating truncation sheets and the Ewald sphere leads to elongated diffraction orders.}
\label{fig:reciprocal_ewald_gratings}
\end{figure*}

To test the lower limits of target sizes in GISAXS, we manufactured a series of grating targets on a single silicon wafer, with each target consisting of 40 grooves of differing line length $l$, forming a grating with pitch $p=\SI{100}{nm}$.
In total, 11 targets were produced in this length series, one ``infinitely" long target with $l=\SI{2500}{\micro\meter}$ and 10 targets with lengths ranging from $l=\SI{50}{\micro\meter}$ down to $\SI{4}{\micro\meter}$.
For all targets, the target width is $\SI{4}{\micro\meter}$, the individual line width is $w=\SI{55}{nm}$ and the nominal line height is $h=\SI{45}{nm}$.
The targets were placed at a distance of \SI{3.04}{mm} from their nearest neighbour to ensure that in conical mounting only one target is hit by the beam.

For the measurements of the very small targets in GISAXS, we need to consider how much of the incoming X-ray beam can interact with the measured target.
Due to the shallow incidence angle, the beam footprint on the sample is enlarged by $1/\sin(\alpha_i)$.
With a beam size of about \SI{0.5x0.5}{mm} and an incidence angle of $\alpha_i=\ang{0.6}$, this yields a beam footprint on the sample of about \SI{0.5x50}{mm}.
The largest target covers an area of \SI{4x2500}{\micro\meter} on the substrate, so only $\approx \num{4e-4}$ of the incident beam interacts with the largest target, and for the smallest target (\SI{4x4}{\micro\meter}), only $\approx \num{6e-7}$ of the beam hits the target.
The scattering from the targets is thus extraordinarily weak and incoherently superimposed with the scattering from the surrounding substrate.
Using suitably long exposure times of $t=\SI{1}{h}$ with the noise-free single photon counting detector, scattering patterns could still be collected.
Additional fitting and subtracting of the diffuse scattering background from the substrate (see supplementary information) allows the scattering patterns of all targets to be obtained.
Measurements for all targets were taken at $E=\SI{6}{keV}$ with an incidence angle of $\alpha_i \approx \SI{0.6}{\degree}$ in conical mounting.

While the scattering from the longest grating (fig. \ref{fig:gisaxs_length_series} a) shows sharp diffraction orders on a semicircle similar to the scattering patterns of infinitely long gratings, shorter gratings show length-dependent changes (fig. \ref{fig:gisaxs_length_series} b) and the shortest grating (fig. \ref{fig:gisaxs_length_series} c) produces a scattering pattern which has lost the circle-like interference pattern almost completely.
For the small ($l \leq \SI{50}{\micro\meter}$) gratings, side lobes above and below the grating diffraction order are visible, and with decreasing length, the diffraction orders as well as the side lobes elongate in the vertical direction and the side lobes move further away from the main peak.
The width of the peaks in the horizontal direction does not change with line length $l$ and is due to the size and divergence of the incoming X-ray beam.

\begin{figure}[t]
\includegraphics[width=\columnwidth]{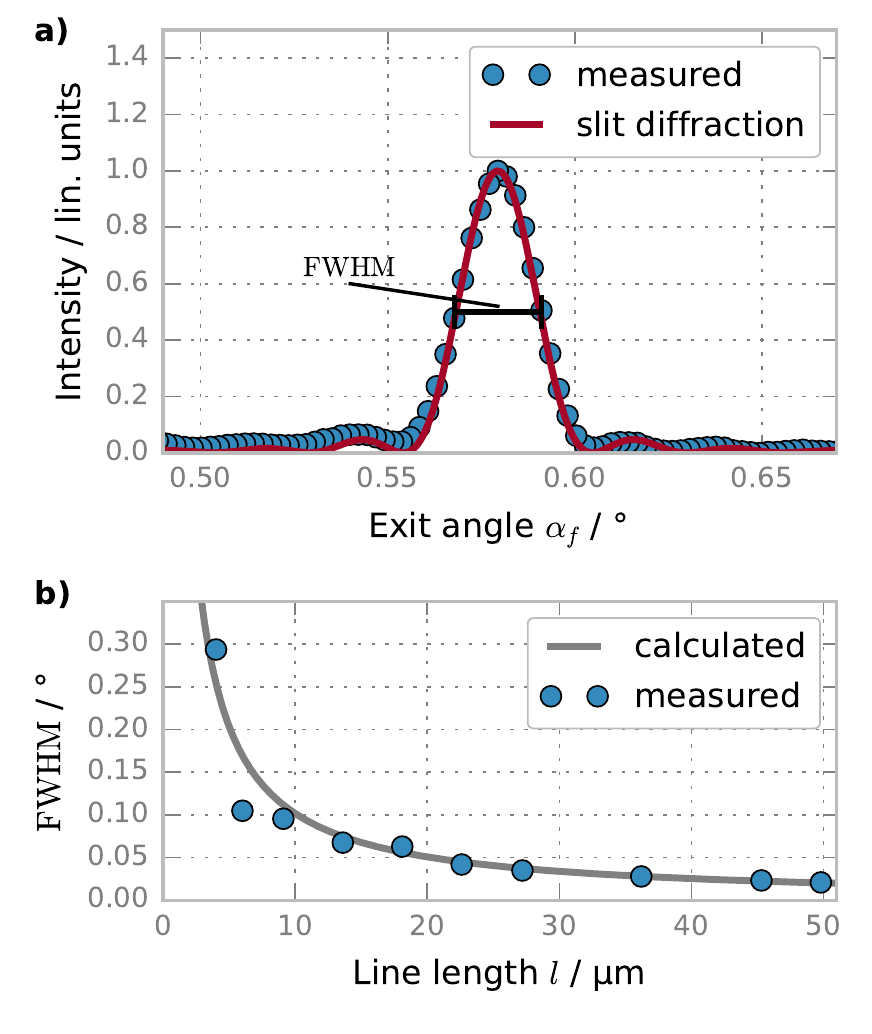}
\caption{Description of scattering of short gratings as single slit diffraction.
\textbf{a)} Cut along the first diffraction order of the scattering of the grating target with $l = \SI{45.3}{\micro\meter}$.
What is shown is the $\mathrm{FWHM}$ extracted from the measured data and the corresponding intensity profile calculated for slit diffraction according to \eqref{eq:single_slit_diffraction}.
\textbf{b)} Comparison between the measured $\mathrm{FWHM}$ extracted from the GISAXS patterns for the length series and the $\mathrm{FWHM}$ calculated from the line length $l$ according to \eqref{eq:fwhm_from_l}.}
\label{fig:gisaxs_length_comparison}
\end{figure}

To explain the changes in the scattering patterns for gratings with finite length, we need to consider the changes in reciprocal space when the grating is finite in the $x$-direction.
The finite length enlarges the grating truncation rods in $k_x$, leading to grating truncation sheets.
The intersection of the grating truncation sheets with the Ewald sphere then leads to elongated diffraction orders (see fig. \ref{fig:reciprocal_ewald_gratings}).
For a quantitative description of the intensity profile along the diffraction orders, we treat the diffraction from short gratings as single-slit diffraction.
The intensity $I$ after diffraction on a single slit is \cite{meschede_wellenoptik_2015}:
\begin{equation}
I = I_0 \left(\mathrm{sinc}\!\left( \frac{s \pi \sin \beta}{\lambda} \right) \right)^2  \label{eq:single_slit_diffraction}
\end{equation}
with the unnormalized cardinal sine function $\mathrm{sinc}(x) = \sin(x)/x$ and the intensity factor $I_0$.
In our case, the effective width of the slit $s$ is the projection of the line length on the incoming beam, $s = l \sin \alpha_i$ and the angle of diffraction $\beta$ is the deviation from the specularly reflected beam, $\beta=\alpha_i - \alpha_f$.
For comparison with the experimental data, we solve \eqref{eq:single_slit_diffraction} numerically for $\beta$ by inserting $I=I_0/2$, which yields:
\begin{equation}
\mathrm{FWHM} = 2 \beta \approx 2 \arcsin \left( \frac{0.443 \lambda}{l \, \sin \alpha_i} \right) \quad , \label{eq:fwhm_from_l}
\end{equation}
for the full width at half maximum of the elongated main peak ($\mathrm{FWHM}$).

\begin{figure}[htb]
\centering
\includegraphics[width=0.9\columnwidth]{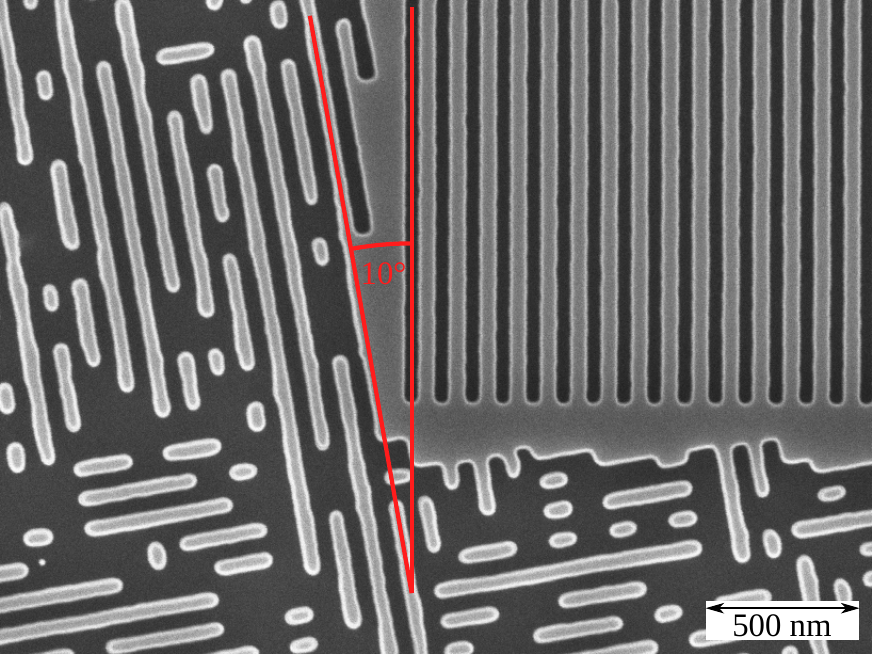}
\caption{Top view SEM image of surrounded field 1, showing the corner of the small grating field (top right) and the surroundings.
Darker areas correspond to etched grooves, lighter areas to mesas.
The orientations of the small grating field and the surroundings, at \ang{10} rotation, are in red.}
\label{fig:sem_sf_edge}
\end{figure}

We have extracted the $\mathrm{FWHM}$ peak width as shown in fig. \ref{fig:gisaxs_length_comparison} a) for all targets in the length series.
The results are shown and compared to the theoretical values from \eqref{eq:fwhm_from_l} in fig. \ref{fig:gisaxs_length_comparison} b).
As can be seen, slit diffraction quantitatively describes the elongation of the main peak and the magnitude of the side lobes due to short line lengths.

\section{Surrounded Small Fields}

\begin{figure*}[tbh]
\includegraphics[width=\textwidth]{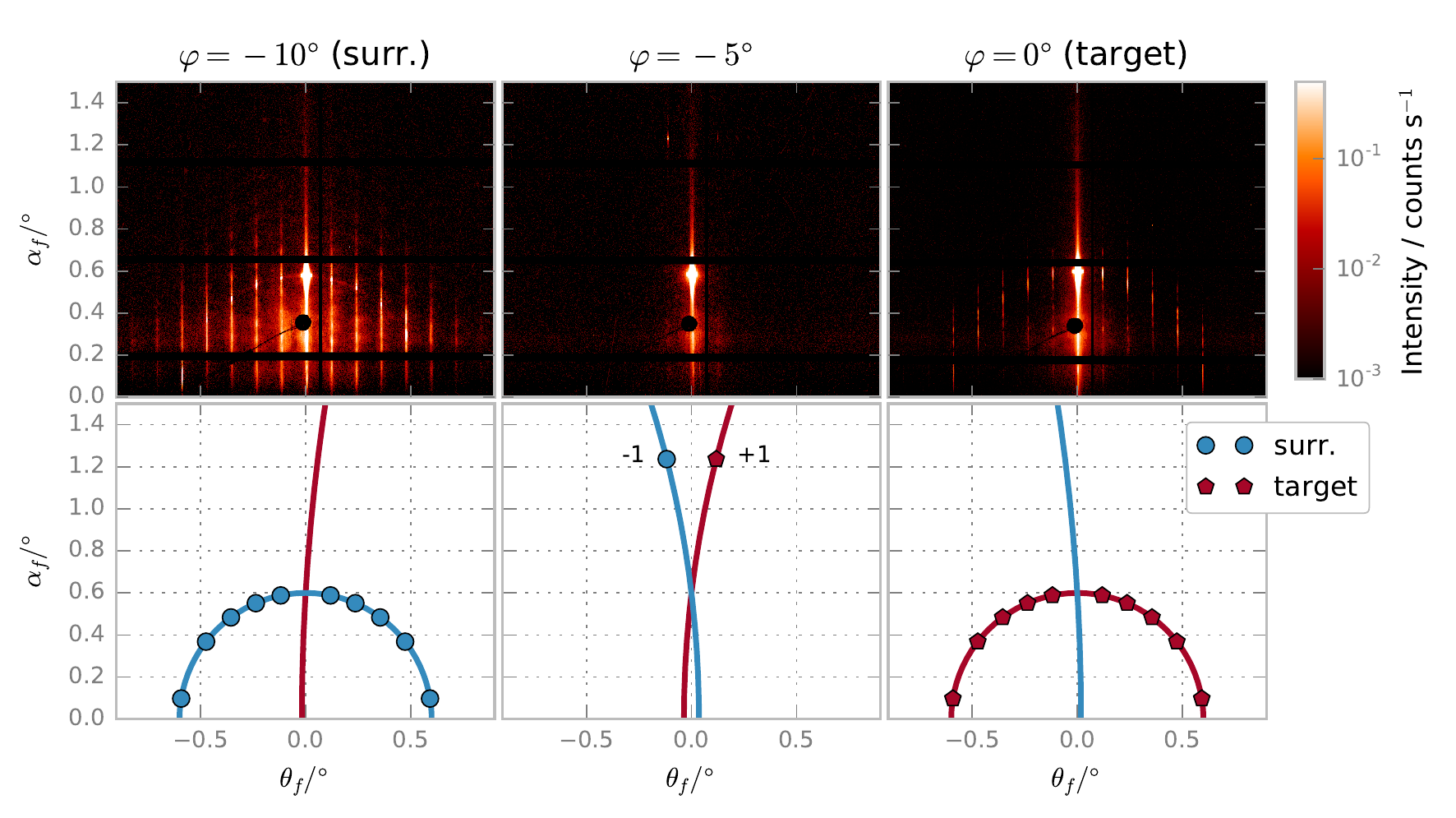}
\caption{GISAXS measurements (upper row) versus theoretical expectation (lower row) of surrounded field 1 at different rotation angles $\varphi$.
At $\varphi = \ang{-10}$ (left) the X-ray beam is oriented along the surrounding structure, showing the scattering orders of the surroundings and a rich diffuse background.
At $\varphi = \ang{-5}$ (middle) the X-ray beam is equally misaligned to the surroundings and the grating target, with only the first diffraction order visible at $\alpha_f\approx \ang{1.2}$ for the surroundings and the target, respectively.
With the X-ray beam aligned to the target ($\varphi = \ang{0}$, right), only the scattering of the target is visible on the detector.
An animated sequence showing scattering patterns from $\varphi = \ang{-10}$ to $\varphi = \ang{0}$ in steps of $\Delta \varphi = \ang{0.1}$ is available in the supplementary information.}
\label{fig:gisaxs_sf45_tilt}
\end{figure*}

In most cases, small targets are not isolated on a blank wafer.
Therefore, it is essential to separate the parasitic scattering of the surroundings from the scattering of the target structure.
One way to separate the scattering of the target and the surroundings if both the target and the surroundings are oriented internally would be a variation of the dominant length scale (for gratings, the pitch $p$) of the target with respect to the surroundings, which would lead to a separation in $\theta_f$.
However, the sensitivity of $\theta_f$ to changes in $p$ is not very high and for surroundings with multiple dominant length scales, it might be difficult to find a suitable $p$ for the target.
Therefore, it is advantageous to rotate the target in the sample plane with respect to the surroundings, which leads to a separation of the scattering in $\alpha_f$.
If the surroundings and the target can be described in good approximation as gratings, this effect can be quantified using \eqref{eq:grating_alpha_f}.

To show a GISAXS measurement of small targets in structured surroundings, we manufactured small grating targets surrounded by ordered but randomized structures, with the grating orientation rotated by \ang{10} with respect to the orientation of the surroundings (see fig. \ref{fig:sem_sf_edge}).
To explore the sensitivity of GISAXS measurements of small grating targets to changes in the target line profile, we manufactured two targets with differing line widths but identical surroundings.
The surroundings measure \SI{100x100}{\micro\meter} and the grating targets at the centre of the surroundings measure \SI{15x15}{\micro\meter}.
For both targets, the surroundings consist of boxes with randomized lengths between \SI{0.2}{\micro\meter} and \SI{3}{\micro\meter}, oriented either in parallel or orthogonally to the standard beam direction.

Both grating targets have a grating pitch of $p=\SI{100}{nm}$ and a nominal line height of $h=\SI{100}{nm}$, but differ in the line width $w$.
For surrounded field 1, the line width is $w=\SI{45}{nm}$ and for surrounded field 2 it is $w=\SI{55}{nm}$.

\begin{SCfigure*}[1.0][htb]
\includegraphics[width=0.65\textwidth]{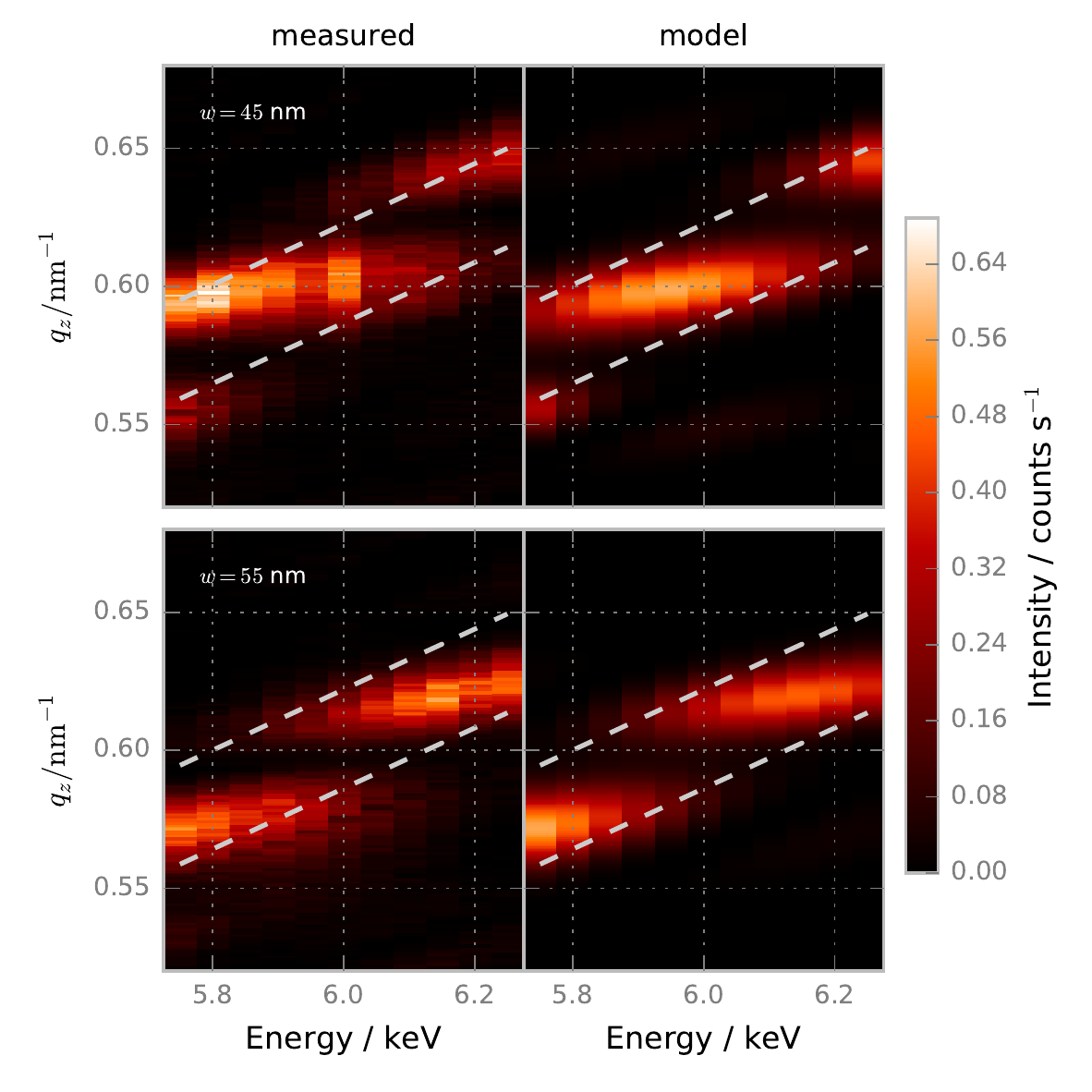}
\caption{Vertical cuts through the second diffraction order of the GISAXS patterns collected at different photon energies (measured data on the left, fitted model on the right).
The measurements for the surrounded field 1 (line width $w=\SI{45}{nm}$, top) and for surrounded field 2 ($w=\SI{55}{nm}$, bottom) are shown.
The dashed lines show the $\mathrm{FWHM}$ of the slit diffraction calculated using \eqref{eq:fwhm_from_l}, which indicates the window of reciprocal space measured at the respective photon energy.
Since detector quantum efficiency and photon flux change with the photon energy, absolute intensities are not comparable between different photon energies.}
\label{fig:gisaxs_sfs_comparison}
\end{SCfigure*}

GISAXS measurements of the surrounded fields were taken with a beam size of \SI{0.1x0.1}{mm}, such that the width of the X-ray beam corresponds to the width of the surroundings.
Measurements were taken at different sample rotations $\varphi$; the results are shown in fig. \ref{fig:gisaxs_sf45_tilt}.
The scattering contributions of the surroundings and the target are well separated and follow the theoretical expectation.
Although the target only covers about \SI{2.3}{\percent} of the structured area, only the scattering of the target is visible on the detector if the beam is aligned with the target.
Due to the high sensitivity of the exit angle $\alpha_f$ to small deviations in the rotation $\varphi$, the grating diffraction orders of the surroundings as well as the diffuse halo originating from the surroundings are suppressed when measuring in target direction, as can be seen by the absence of scattering originating from the surroundings if the beam is aligned just between the target and the surroundings ($\varphi=\ang{-5}$).

We measured target GISAXS patterns ($\varphi=\ang{0}$) at photon energies from $E=\SI{5750}{eV}$ up to $E=\SI{6250}{eV}$ for both surrounded field 1 (line width $w=\SI{45}{nm}$) and surrounded field 2 ($w=\SI{55}{nm}$).
Vertical cuts through the second diffraction order (at $q_y=\SI[per-mode=reciprocal]{0.126}{\per\nm}$) for both targets and all energies are shown in fig. \ref{fig:gisaxs_sfs_comparison}.
The measurements can be understood in terms of the reciprocal space construction.
Within this framework, changing the photon energy alters the radius of the Ewald sphere and consequently the position of the intersection between the Ewald sphere and the grating truncation sheets.
Effectively, we measure a different part of the grating truncation sheets at each energy, explaining why the cuts show zero intensity outside of this window into reciprocal space.
The intensity profile within the measured window can be explained using a model composed of Gaussian peaks at constant positions in $q_z$ multiplied with the energy-dependent slit diffraction according to \eqref{eq:single_slit_diffraction}.
Figure \ref{fig:gisaxs_sfs_comparison} shows models fitted to the data using the known length $l=\SI{15}{\um}$ for the slit diffraction and three or respectively two Gaussian peaks for $w=\SI{45}{nm}$ and $w = \SI{55}{nm}$.
While relative intensities are not accurately represented, the models describe peak positions very well, showing that the intensity profile within the measured window is explained by target features in the $z$-direction.
The distance between the peaks is about $\Delta q_z = \SI[per-mode=reciprocal]{0.5}{\per\nm}$, which roughly corresponds to the nominal line height of $h=\SI{100}{\nm}$ in real space along $z$.
Comparing the measurements for the two targets with different line widths, $\Delta q_z$ does not change significantly, but the position of the peaks is shifted.
From previous studies \cite{suh_characterization_2016,soltwisch_reconstructing_2017} on practically infinitely long gratings, it is known that the intensity of the non-elongated grating diffraction orders depends on the exact line profile.
We therefore attribute the changes in position and relative intensity of the observed peaks within the elongated diffraction orders to the differences in line profile, mainly the differing line widths.

\section{Conclusions}

We have shown that even with millimetre-sized beams, which are available from many synchrotron and lab-based X-ray sources, micrometre-sized targets can be measured.
The minimum target sizes which were investigated are an order of magnitude smaller than the smallest micro-beam footprints which have been used in GISAXS experiments so far \cite{roth_situ_2007}.
The challenge in the measurements is separating the scattering signal of the target from the scattering of its surroundings.
While this separation is easily done for trivial surroundings like a bare substrate, it becomes more challenging if the scattering of structured surroundings and the target overlap.
We managed to separate the scattering of periodic targets in nanostructured surroundings if the targets were rotated with respect to the predominant direction of the surroundings.

The presented formulas for single-slit diffraction describe the elongation of grating diffraction orders and the appearance of side lobes when going from effectively infinite to short targets.
The comparison of the scattering of two small grating targets with different line widths shows that GISAXS measurements of small targets are sensitive to the grating line profile.
For infinitely long grating lines, previous studies using the DWBA \cite{suh_characterization_2016} or a Maxwell solver \cite{soltwisch_reconstructing_2017} reduced the calculations of GISAXS measurements to two dimensions and were then able to reconstruct the full line profile.
As short lines are inherently three-dimensional, further research is needed to extend these methods to the reconstruction of line profiles of short grating targets.

Using the techniques described in this paper, it is possible to employ GISAXS with its distinct advantages for applications such as characterization of metrology fields in the semiconductor industry where up to now it has been considered impossible to use GISAXS due to the large beam footprint.

\section{Acknowledgements}
The authors wish to thank Analía Fernández Herrero and Anton Haase for their helpful discussions and Levent Cibik, Stefanie Langner and Swenja Schreiber for their support during experiments.

V. Soltwisch and M. Pflüger have applied for a German patent claiming inventions partly described in this paper \cite{soltwisch_verfahren_2017}.


\bibliography{zotero}

\end{document}


\date{}
\title{Supporting Information}
\subtitle{Grazing Incidence Small Angle X-Ray Scattering (GISAXS) on Small Targets Using Large Beams}
\author{Mika Pflüger \and Victor Soltwisch \and Jürgen Probst \and Frank Scholze \and Michael Krumrey}
\maketitle

\section{Background Correction}

\begin{figure}[hb]
\includegraphics{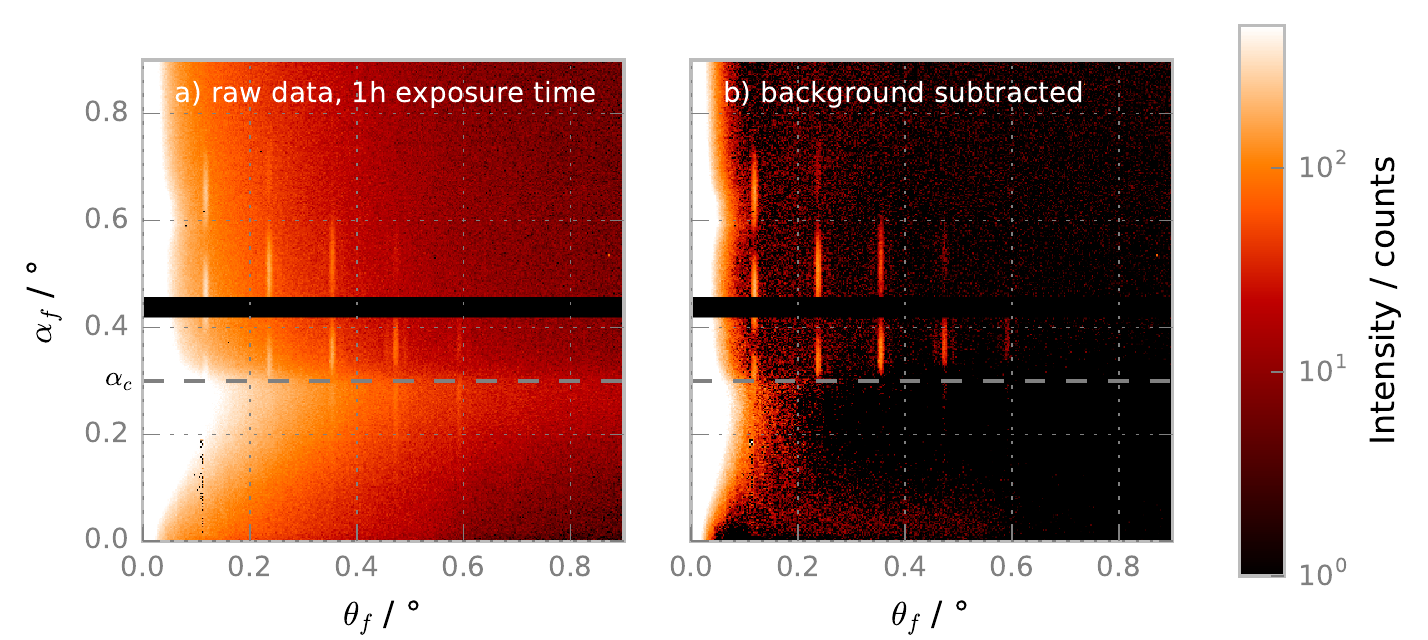}
\caption{GISAXS scattering of smallest target, \textbf{a)} raw data \textbf{b)} after background subtraction.
The background subtraction works well above the critical angle of the substrate $\alpha_c$, but fails below $\alpha_c$.}
\label{fig:gisaxs_4x4}
\end{figure}
In order to extract the scattering of the targets only, the background $B$ was fitted for each measurement, assuming that the background $B$ can be factorized to $B(\alpha_f, \theta_f) = A(\alpha_f) \cdot T(\theta_f)$.
For the function $A(\alpha_f)$, a smooth B-spline approximation of degree 2 was used to closely follow the scattering of the background around the critical angle of total external reflection $\alpha_c$ of the substrate.
In order to only fit the substrate contribution, a cut along $\alpha_f$ was taken between the first and second grating diffraction orders.
For the function $T(\theta_f)$, a polynomial of degree 4 was fitted to a cut along $\theta_f$ at $\alpha_f > \SI{0.8}{\degree}$, i.e. above the sample scattering features. The resulting smooth background was subtracted from the GISAXS measurement, yielding the scattering from the target only (fig. \ref{fig:gisaxs_4x4}).

\section{Position of Grating Diffraction Orders in GISAXS in Sample Coordinates}

\subsection{Coordinate System and Ewald Sphere}

We use a coordinate system where the $x$-$y$-plane is the sample plane, with the $x$-axis the intersection of the scattering plane with the sample plane and the $y$-axis perpendicular to the $x$-axis. The $z$-axis is the normal of the sample plane. The $k$-space is the reciprocal of the real space, with the corresponding axes in the same direction as the real axes. In this space, the wavevectors of the incoming beam $\bm k_i$ and the scattered beam $\bm k_f$ are
\begin{align}
\bm k_i &= k_0 \left(\begin{matrix}
\cos\alpha_i \\
0 \\
-\sin\alpha_i
\end{matrix}\right) \\
\bm k_f &= k_0 \left(\begin{matrix}
\cos\alpha_f \, \cos\theta_f \\
\cos\alpha_f \, \sin\theta_f \\
\sin\alpha_f
\end{matrix}\right) \\
k_0 &= \abs{\bm k_i} = \abs{\bm k_f} = \frac{2\pi}{\lambda} \label{eq_k0}
\end{align}
with the incident angle $\alpha_i$, the angle between the sample plane and the scattered beam $\alpha_f$ and the angle between the projection of the scattered beam on the sample plane and the $x$-axis $\theta_f$ as well as the incident wavelength $\lambda$.

We define the scattering vector $\bm q = \bm k_f - \bm k_i$, which expressed in angle coordinates is
\begin{align}
\bm q &= k_0 \left(\begin{matrix}
\cos\alpha_f \, \cos\theta_f - \cos\alpha_i\\
\cos\alpha_f \, \sin\theta_f \\
\sin\alpha_f + \sin\alpha_i
\end{matrix}\right), \label{eq_q_angle}
\end{align}
together with \eqref{eq_k0} we can write the equation for the Ewald sphere of elastic scattering
\begin{align}
k_0 &= \abs{\bm k_f} = \abs{\bm q + \bm k_i} \\
\Rightarrow k_0^2 &= \abs{\bm q + \bm k_i}^2 = (q_x+k_{i,x})^2 + (q_y+k_{i,y})^2 + (q_z+k_{i,z})^2 \nonumber \\
 &= (q_x + k_0 \cos \alpha_i)^2 + q_y^2 + (q_z - k_0 \sin \alpha_i)^2 \quad . \label{eq_ewald}
\end{align}

\subsection{Perfectly Aligned Grating}
The perfectly aligned grating has infinite grating lines parallel to the $x$-axis, which lie in the sample plane and are separated by the pitch $p$. The reciprocal space representation of the perfectly aligned grating comprises grating truncation rods (GTR), which are parallel to the $q_z$-axis in the $q_z$-$q_y$-plane and separated by $2\pi/p$ in $q_y$:
\begin{align}
q_x &= 0 \label{eq_agtr_q_x} \\
q_y &= n\, 2\pi/p = k_0\, n \lambda/p \label{eq_agtr_q_y}
\end{align}
with the grating diffraction order $n \in \ZZ$.
The intersection of the Ewald sphere \eqref{eq_ewald} with the GTR yields
\begin{align}
k_0^2 &= (0 + k_0 \cos \alpha_i)^2 + (n\, k_0\, \lambda/p)^2 + (q_z - k_0 \sin \alpha_i)^2 \\
& \textrm{solving for $q_z$} \nonumber \\
(q_z - k_0 \sin \alpha_i)^2 &= k_0^2 (1 - \cos^2 \alpha_i) - (n\, k_0\, \lambda/p)^2 \nonumber \\
                            &= k_0^2 \left(\sin^2 \alpha_i - (n\, \lambda/p)^2\right) \\
\Rightarrow q_z &= k_0 \left(\sin\alpha_i \pm \sqrt{\sin^2\alpha_i - (n\, \lambda/p)^2} \right) \\
& \textrm{discarding the solution with the minus as it} \nonumber \\
& \textrm{corresponds to reflections below the sample horizon} \nonumber \\
q_z &= k_0 \left(\sin \alpha_i + \sqrt{\sin^2 \alpha_i -(n \lambda / p)^2}\right) \quad . \label{eq_agtr_q_z}
\end{align}
To summarize:
\begin{align}
\bm q_{\textrm{grating, aligned}} &= k_0 \left(\begin{matrix}
0 \\
n \lambda / p \\
\sin \alpha_i + \sqrt{\sin^2 \alpha_i - (n\lambda / p)^2}
\end{matrix}\right) \quad .
\end{align}

To express the scattering in angle coordinates, we use \eqref{eq_q_angle}, \eqref{eq_agtr_q_x}, \eqref{eq_agtr_q_y} and \eqref{eq_agtr_q_z}, giving
\begin{align}
q_z &: & \sin \alpha_f + \sin \alpha_i &= \sin \alpha_i \left(1 + \sqrt{1-\left(\frac{n\lambda}{p \sin\alpha_i}\right)^2}\right) \nonumber\\
& &\Rightarrow \alpha_f &= \arcsin\left(
 \sqrt{
  \sin^2\alpha_i  - 
  \left(
   \frac{n\lambda}{p}
  \right)^2}
\right) \\
q_y &: & \cos \alpha_f \sin \theta_f &= n \lambda / p \nonumber \\
    &  &\Rightarrow  \sin \theta_f &= \frac{n \lambda / p}{\cos \alpha_f} \\
q_x &: & \cos \alpha_f \cos \theta_f - \cos \alpha_i &= 0 \nonumber \\
    &  &\Rightarrow \cos \theta_f &= \frac{\cos \alpha_i}{\cos \alpha_f} \\
\frac{q_y}{q_x} &: & \tan \theta_f = \frac{\sin \theta_f}{\cos \theta_f} &= \frac{n\lambda/p}{\cos \alpha_f} \frac{\cos \alpha_f}{\cos \alpha_i} \nonumber \\
    &  &\Rightarrow \theta_f &= \arctan\left(\frac{n \lambda}{p \cos \alpha_i}\right)
\end{align}

\subsection{Misaligned Grating}
For the misaligned grating, the grating lines are rotated around the $z$-axis by $\varphi$, and thus the GTRs are also rotated around the $k_z$-axis by $\varphi$, giving the conditions
\begin{align}
q_x &= k_0 \sin \varphi \,n \lambda/p \label{eq_mgtr_q_x} \\
q_y &= k_0 \cos \varphi \,n \lambda/p \quad . \label{eq_mgtr_q_y}
\end{align}
The intersection with the Ewald sphere \eqref{eq_ewald} now yields
\begin{align}
k_0^2 &= (k_0 \sin \varphi \,n \lambda/p + k_0 \cos \alpha_i)^2 + (k_0 \cos \varphi \,n \lambda/p)^2 + (q_z - k_0 \sin \alpha_i)^2 \nonumber \\
 &= k_0^2 \left( (\sin^2 \varphi + \cos^2 \varphi) (n \lambda/p)^2 + 2 \sin \varphi \cos \alpha_i \,n \lambda/p + \cos^2 \alpha_i \, \right) + (q_z - k_0 \sin \alpha_i)^2
\end{align}
solving for $q_z$
\begin{align}
 (q_z - k_0 \sin \alpha_i)^2 &= k_0^2 \left(1 - \cos^2 \alpha_i - (n \lambda/p)^2 - 2 \sin \varphi \cos \alpha_i \,n \lambda/p \right) \nonumber \\
  &= k_0^2 \left(\sin^2 \alpha_i - (n \lambda/p)^2 - 2 \sin \varphi \cos \alpha_i \,n \lambda/p \right) \\
\Rightarrow q_z &= k_0 \left(\sin \alpha_i \pm \sqrt{\sin^2 \alpha_i - (n \lambda/p)^2 - 2 \sin \varphi \cos \alpha_i \,n \lambda/p} \right) \\
& \textrm{discarding the solution with the minus as it} \nonumber \\
& \textrm{corresponds to reflections below the sample horizon} \nonumber \\
q_z &= k_0 \left(\sin \alpha_i + \sqrt{\sin^2 \alpha_i - (n \lambda/p)^2 - 2 \sin \varphi \cos \alpha_i \,n \lambda/p} \right) \quad . \label{eq_mgtr_q_z}
\end{align}
To summarize:
\begin{align}
\bm q_{\textrm{grating}} &= k_0 \left(\begin{matrix}
\sin \varphi \,n \lambda/p \\
\cos \varphi \,n \lambda/p \\
\sin \alpha_i + \sqrt{\sin^2 \alpha_i - (n \lambda/p)^2 - 2 \sin \varphi \cos \alpha_i \,n \lambda/p}
\end{matrix}\right) \quad .
\end{align}

To express the scattering in angle coordinates, we use \eqref{eq_q_angle}, \eqref{eq_mgtr_q_x}, \eqref{eq_mgtr_q_y} and \eqref{eq_mgtr_q_z}, giving
\begin{align}
q_z &: & \sin \alpha_f + \sin \alpha_i &= \sin \alpha_i + \sqrt{\sin^2 \alpha_i - (n \lambda/p)^2 - 2 \sin \varphi \cos \alpha_i \,n \lambda/p} \nonumber \\
& & \Rightarrow \alpha_f &= \arcsin\left(\sqrt{\sin^2 \alpha_i - (n \lambda/p)^2 - 2 \sin \varphi \cos \alpha_i \,n \lambda/p}\right) \\
q_y &: & \cos \alpha_f \sin \theta_f &= \cos \varphi \,n \lambda/p \nonumber \\
& & \Rightarrow \sin \theta_f &= \frac{\cos \varphi \,n \lambda/p}{\cos \alpha_f} \\
q_x &: & \cos \alpha_f \cos \theta_f - \cos \alpha_i &= \sin \varphi \, n \lambda / p \nonumber \\
& & \Rightarrow \cos \theta_f &= \frac{\sin \varphi \, n \lambda / p + \cos \alpha_i}{\cos \alpha_f} \\
\frac{q_y}{q_x} &: & \tan \theta_f = \frac{\sin \theta_f}{\cos \theta_f} &= \frac{\cos \varphi \,n \lambda/p}{\cos \alpha_f} \frac{\cos \alpha_f}{\sin \varphi \, n \lambda / p + \cos \alpha_i}  \nonumber \\
& & \Rightarrow \theta_f &= \arctan \left( \frac{\cos \varphi \, n \lambda /p}{\sin \varphi \, n \lambda / p + \cos \alpha_i}  \right) \quad .
\end{align}